\documentclass{agujournal2018}
\usepackage{apacite}
\usepackage{url} 

\journalname{JGR-Space Physics}

\begin{document}

%
%


\title{Snakes on a Spaceship -- An Overview of Python in Heliophysics}

%
%




\authors{
A.G.~Burrell\affil{1,2}, 
A.~Halford\affil{3}, 
J.~Klenzing\affil{4},  
R.A.~Stoneback\affil{2},
S.~K. Morley\affil{5},
A.M.~Annex\affil{6},
K.M.~Laundal\affil{7},
A.C.~Kellerman\affil{8},
D.~Stansby\affil{9},
J.~Ma\affil{10}
}


\affiliation{1}{Space Science Division, U.S. Naval Research Laboratory, Washington, DC, USA}
\affiliation{2}{William B. Hanson Center for Space Sciences, The University of Texas at Dallas, Richardson, TX, USA}
\affiliation{3}{Space Sciences Department, The Aerospace Corporation, Chantilly, VA}
\affiliation{4}{ITM Physics Laboratory / Code 675, Goddard Space Flight Center Greenbelt, MD, 20771}
\affiliation{5}{Space Science and Applications, Los Alamos National Laboratory, Los Alamos, NM 87545, USA}
\affiliation{6}{Department of Earth and Planetary Sciences, Johns Hopkins University, Baltimore, MD, 21218}
\affiliation{7}{Birkeland Centre for Space Science, University in Bergen, Norway}
\affiliation{8}{Department of Earth Planetary and Space Sciences, University of California, Los Angeles, CA}
\affiliation{9}{Imperial College London, London, SW7 2AZ, UK}
\affiliation{10}{Facebook AI Research, Menlo Park, CA 94025}





\correspondingauthor{A.G. Burrell}{angeline.burrell@nrl.navy.mil}




\begin{keypoints}
\item A wealth of Python analysis tools exist to support space physics research
\item Open source tools aid reproducible science
\item Collaborative programming is essential for analysis tools to keep pace with scientific progress
\end{keypoints}

%
%


\begin{abstract}
Computational analysis has become ubiquitous within the heliophysics community.
However, community standards for peer-review of codes and analysis have lagged behind these developments. 
This absence has contributed to the reproducibility crisis, where inadequate analysis descriptions and loss of scientific data have made scientific studies difficult or impossible to replicate.

The heliophysics community has responded to this challenge by expressing a desire for a more open, collaborative set of analysis tools.  
This article summarizes the current state of these efforts and presents an overview of many of the existing Python heliophysics tools.
It also outlines the challenges facing community members who are working towards the goal of an open, collaborative, Python heliophysics toolkit and presents guidelines that can ease the transition from individualistic data analysis practices to an accountable, communalistic environment.  

{\bf Plain Language Summary}
As computers have become more powerful and better at solving complex mathematical equations, space scientists have relied more and more on computational tools. 
Community standards for peer-review of computer codes and similar analysis tools have not kept pace with the development of these technologies.
This lag in community accountability has contributed to a crisis in the scientific literature, where it has been hard or even impossible to verify past studies.

Space scientists have responded to this challenge with a desire for open, shared analysis tools.  
This article summarizes the current state of these efforts and presents an overview of many of the existing Python analysis tools used in space physics.
It also outlines the challenges facing scientists who are working towards the goal of an open, shared, Python space science toolkit and presents guidelines that can ease the transition from private to public data analysis practices. 

\end{abstract}

%
%

\section{Introduction}
The proliferation of observed and modeled data within the field of heliophysics, a field that encompasses solar, magnetospheric, and  upper atmospheric studies within the solar system, has vastly expanded the possibilities for science investigation.  These ever expanding archives (e.g., the Coupling, Energetics and Dynamics of Atmospheric Regions (CEDAR) Madrigal database, the National Aeronautics and Space Administration (NASA) Coordinated Data Analysis Web (CDAWeb), and the Near Earth Space data infrastructure for e-science (ESPAS)) are accessible and searchable, but (appropriately) do not provide many tools for identifying specific case studies or performing data analysis.  This absence has contributed to the reproducibility crisis, where the results of scientific studies have been shown to be difficult or even impossible to replicate \citep{peng:2011, gil:2016}.  Additionally, as the heliophysics community moves toward solving more complex interdisciplinary problems through data science techniques, a common infrastructure capable of handling diverse data sets is required \citep{McGranaghan:2017kd}.

The challenges presented by large, distinct data sets and unreproducible results may be surmounted using currently available tools and techniques.  Other scientific communities have tackled these challenges, providing an example for the heliophysics community to follow.  For example, the Incorporated Research Institutions for Seismology (IRIS), provides services that include batch data downloads and searchable lists of institution and community developed software \citep{iris}.  This article presents a framework for creating, maintaining, and sharing these tools within the space physics community.  It begins in section~\ref{benefits.sec} by discussing the benefits of open source tools in general, and Python \citep{python} in particular, to the scientific community.  Section~\ref{packages.sec} then presents a summary of currently available heliophysics Python packages.  Finally, a framework for future community development is presented in section~\ref{framework.sec}.  The desire for an overarching framework (inspired by successful projects in other disciplines, such as AstroPy \citep{astropy:2013}) has been repeatedly expressed in space physics community meetings and workshops \citep<e.g.,>{agu:2017, cedar:2017}.

\section{Upholding Mertonian norms}
\label{benefits.sec}

The modern, Western scientific ethos can be described by the Mertonian norms of universalism, communalism, disinterestedness, and organized skepticism \citep{merton:1957}.  
Universalism speaks to the expectation that scientific results will be evaluated on their own merit.  
Communalism asserts that scientific knowledge belongs to the entire community, not an individual or single institution.  
Disinterestedness espouses the rejection of personal gain, in terms of both prestige and income.  
Organized skepticism requires scientists to be critical of new and old ideas presented by themselves and others in the community.  
These norms, intended to characterize the social aspects of scientific culture, show a strong subscription across disciplines.  
The vast majority of scientists believe, to a great extent, that these norms should be upheld \citep{anderson:2010}.
All of these norms support the adoption of the open source philosophy, which refers to things that have been made publicly accessible for people to use, modify, and share.  

Making science and scientific analysis open source has a wide range of benefits.  
It encourages organized skepticism and addresses the reproducibility crisis \citep{gil:2016}, since all community members would have all of the necessary information available to independently evaluate and build upon past work \citep{peng:2011}.
With proper citation, it brings scientific work that is currently being performed behind the scenes into the open where it can be appropriately evaluated and recognized, reducing opportunities for methodological plagiarism.
It also reduces gatekeeping (upholding communalism and universalism), since data and analysis tools would not be hidden behind a paywall.  

The reduction of gatekeeping has both ethical and practical benefits, as reduced gatekeeping has been shown to increase diversity within the field~\citep{Murray2018}.
%
%
%
%
Although the benefits of a diverse scientific community are obvious, research has shown that diverse working groups produce higher quality and better cited research \citep{Jehn:1999,Nielsen:2017, AlShebli:2018}. 
These practical benefits also improve the working lives of individual researcher, leading to improved well-being and job satisfaction \citep{kanter2008, Choi2017}.
%
%
%
%
%
%

%
Even though open science both fits with the Western scientific ethos and has been shown to help researchers succeed through increased acknowledgement for traditionally unacknowledged work, increased citations, media attention, funding opportunities, and potential collaborations \citep{McKiernan:2016os}, there is still considerable resistance across the community towards adopting these practices.  
Opponents of open source practices in particular (as open source practices are the focus of this paper) give a range of reasons to keep scientific analysis and data private, including plagiarism, misuse of open science products, risks to career advancement, previous lack of publication opportunities, giving up an edge in funding opportunities, lack of funding for open source software in space sciences, the possibility of producing a new set of gatekeepers with additional requirements for scientific productivity, and the capturing of academic labor output by commercial interests \citep<e.g.,>{Tyfield2013,Longo:2016md,Lancaster:2016}.
%
%
%
%
%
%
%
%
There is also a trend for many researchers to agree with open code policies in principle, but not in practice \citep{Shamir2013}.  
The adoption of open source software practices can be hampered by a lack of familiarity with available tools and by concerns about short term productivity.
Although the impacts of the open source movement on research and individual scientists are ongoing, its early impacts are positive \citep[and references therein]{McKiernan:2016os}.
Following the steps and procedures outlined section~\ref{framework.sec} will help the space physics community avoid many of the potential negative effects of open source science.

Open source science is most easily disseminated when scientists program their tools in an open source language.
%
%
%
%
While there are many options for open source languages, this paper focuses on the applicability of Python to space science.
Python is a popular option for building scientific data analysis tools, because it is a community driven, open source language, with a broad spectrum of well developed packages.
%
%
%
%
%
%
%
%
%
%
Two features make Python stand out as a language: the relative maturity of Python for scientific applications and the widespread use of Python outside of academia. 

Scientific programming in Python typically builds off of the NumPy \citep{numpy} and SciPy \citep{scipy} libraries, which form a mature foundation for scientific applications. 
A rich ecosystem of software packages that build on the `scientific stack' of Python (NumPy and SciPy) are available; this paper describes the current heliophysics ecosystem. 
In addition, legacy codes in languages like Fortran (FORmula TRANslation, \citet{backus}) and C \cite{Ritchie:1996} are easily wrapped in Python.  
Many of the models discussed in Section~\ref{model.sec} utilize this functionality to make empirical models more widely available to the community.  
Python is a high-level language that natively handles many of the required computational tasks (such as memory management and compilation) without requiring any instruction from the developer.
This lowers the barrier for new scientific programmers and allows scientists to focus on their algorithms and analysis rather than the details of computer programming.
Finally, SciPy provides the functionality to read proprietary data sets, including those written by the Interactive Data Language (IDL, \citet{idl}) and Matrix Laboratory (MATLAB, \citet{matlab}), allowing for a smooth transition in working environments that use legacy codes and datasets.

The widespread use of Python outside of the immediate heliophysics community has a number of advantages.  
Commonly-used techniques such as signal processing have a large support network, allowing the heliophysics community to benefit from outside expertise.
It ensures continued language development and support for new technology, since its popularity throughout the global programming community provides the critical mass needed for such investment.
Python has been broadly adopted as a teaching language \citep{Fangohr2004}, allowing the heliophysics community to build on the budding expertise of computationally literate students. 
 Additionally, the use of Python in our community provides students with transferable skills for industry careers \citep{TIOBE,PYPL}.

\section{Overview of Current Packages}
\label{packages.sec}

\begin{figure}[htbp]
\begin{center}
\includegraphics[width=\textwidth]{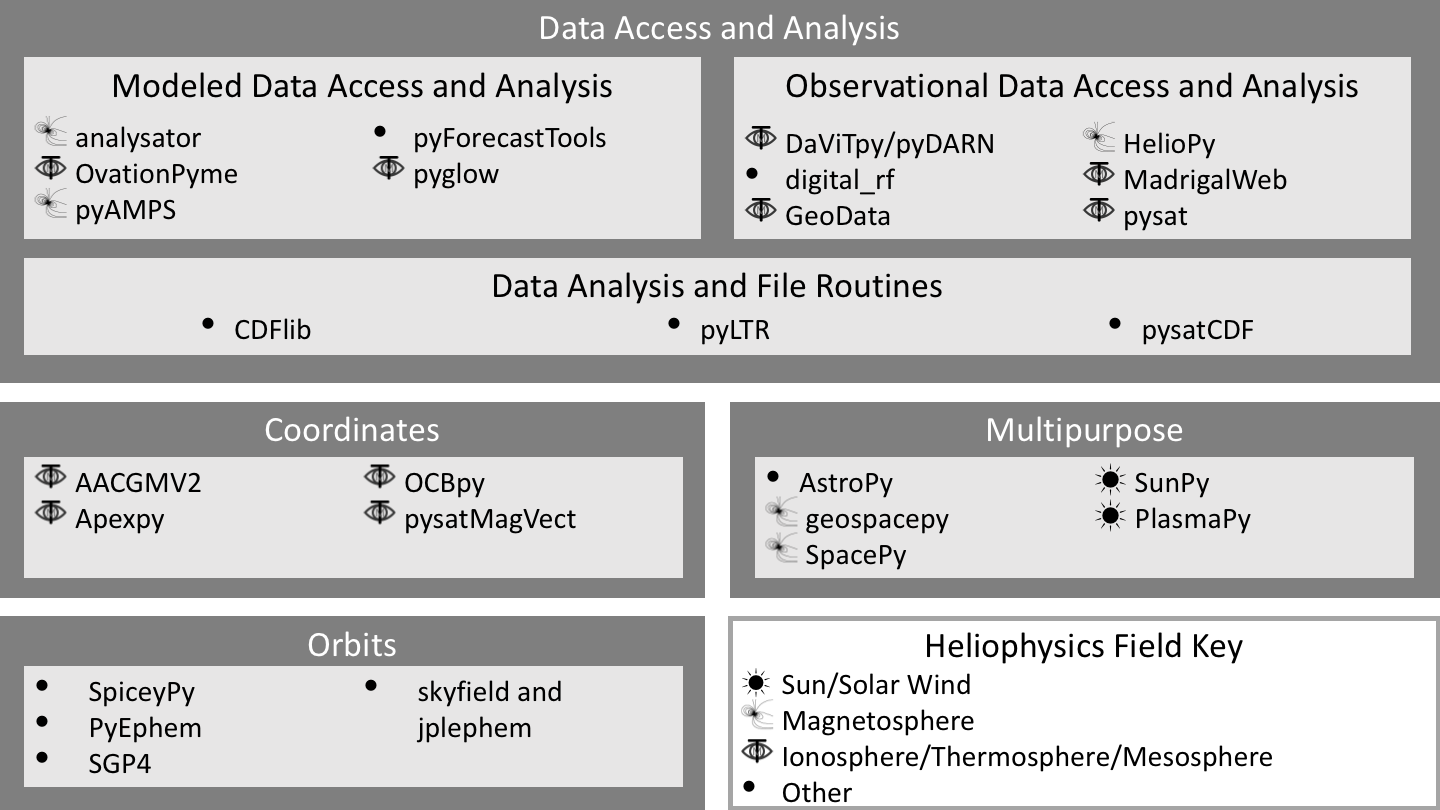}
\caption{Community developed heliophysics packages, grouped by field and purpose.}

\label{map.fig}
\end{center}
\end{figure}

Heliophysics is a diverse research community, with research interests reaching from the sun to the lower thermosphere of solar planets and methods encompassing active experiments, ground and space-based observations, modeling, and theory.  An enterprise this vast requires a wide range of tools.  Figure~\ref{map.fig} outlines the types of Python packages currently available and their regions of specialization, while Table~\ref{packages.tab} supplies their licensing information, package location, and the section in appendix~\ref{appendix.sec} that contains a description of each package.  Although not a complete list of heliophysics Python packages, this figure shows the community written packages commonly used at the time of publication, as determined through an international survey designed to gauge community involvement in collaborative space physics python projects \citep{Burrell:2018}.  This survey was distributed to six space physics mailing lists and received 223 responses (195 complete and 28 partial) from heliophysicists of all career stages across 28 countries and 6 continents.  These project descriptions are also limited to include only those that are all free and open-source software (FOSS). 

\begin{table}
\begin{center}
\caption{License, description section, and location of alphabetically ordered community developed heliophysics packages.  License acronyms are defined in section~\ref{access.sec}.}
\label{packages.tab}
\begin{tabular}{llll}
Name & License & Section & Location \\
\hline 
AACGMV2 &  MIT & \ref{aacgmv2.sec} & \url{https://github.com/aburrell/aacgmv2} \\
analysator &  GPL-2 & \ref{analysator.sec} & \url{https://github.com/fmihpc/analysator} \\
apexpy & MIT & \ref{apexpy.sec} & \url{https://github.com/aburrell/apexpy} \\
Astropy &  BSD & \ref{astropy.sec} & \url{http://www.astropy.org} \\
CDFlib & MIT & \ref{cdflib.sec} & \url{https://github.com/MAVENSDC/cdflib} \\
DaViTPy/pyDARN & GPL-3.0 & \ref{pydarn.sec} & \url{https://github.com/vtsuperdarn/davitpy} \\
digital\_rf &  BSD & \ref{digitalrf.sec} & \url{https://github.com/MITHaystack/digital_rf} \\
GeoData & MIT & \ref{geodata.sec} & \url{https://github.com/jswoboda/GeoData} \\
geospacepy & MIT & \ref{geospacepy.sec} & \url{https://github.com/lkilcommons/geospacepy-lite} \\
HelioPy & GPL-3.0 & \ref{heliopy.sec} & \url{http://docs.heliopy.org} \\
jplephem & MIT & \ref{skyfield.sec} & \url{https://pypi.org/project/jplephem} \\
MadrigalWeb & MIT & \ref{madrigalweb.sec} & \url{http://cedar.openmadrigal.org} \\
OCBpy & BSD-3-Clause & \ref{ocbpy.sec} & \url{https://github.com/aburrell/ocbpy} \\
OvationPyme & LGPL-3.0 & \ref{ovationpyme.sec} & \url{https://github.com/lkilcommons/OvationPyme} \\
PlasmaPy & BSD+Patent & \ref{plasmapy.sec} & \url{http://www.plasmapy.org} \\
pyAMPS & MIT & \ref{pyamps.sec} & \url{https://github.com/klaundal/pyAMPS} \\
PyEphem & LGPL & \ref{pyephem.sec} & \url{http://rhodesmill.org/pyephem/index.html} \\
pyForecastTools & BSD-3-Clause & \ref{pyforecasttools.sec} & \url{https://github.com/drsteve/PyForecastTools} \\
pyglow & MIT & \ref{pyglow.sec} & \url{https://github.com/timduly4/pyglow} \\
pyLTR & BSD-3-Clause & \ref{pyltr.sec} & \url{https://github.com/jma127/pyltr} \\
pysat & BSD-3-Clause & \ref{pysat.sec} & \url{https://github.com/rstoneback/pysat} \\
pysatCDF & BSD-3-Clause & \ref{pysatcdf.sec} & \url{https://github.com/rstoneback/pysatcdf} \\
pysatMagVect & BSD-3-Clause & \ref{pysatMagVect.sec} & \url{https://github.com/rstoneback/pysatmagvect} \\
SGP4 & MIT & \ref{sgp4.sec} & \url{https://pypi.org/project/sgp4} \\
skyfield & MIT & \ref{skyfield.sec} &  \url{http://rhodesmill.org/skyfield} \\
SpacePy & PSF & \ref{spacepy.sec} & \url{https://github.com/spacepy/spacepy} \\
SpiceyPy &  MIT & \ref{spiceypy.sec} & \url{https://github.com/AndrewAnnex/SpiceyPy} \\
SunPy & MIT & \ref{sunpy.sec} & \url{https://sunpy.org} \\
\hline
\end{tabular}
\end{center}
\end{table}
%
%
%
%
At the time of publication, software development within the space physics community is largely an individual effort.  For example, the survey results reported that 92.9\% of respondents wrote their own analysis code and only 18\% contributed to community packages.  This has lead to some overlap of functionality in the community written packages. 

Common areas of overlap include file handling routines, time handling utilities, legacy model implementations, and coordinate transformations.
For example, there are three different packages that load NASA common data format (CDF) files and three different packages that use the International Geomagnetic Reference Magnetic Field (IGRF; \citet{IGRF2015}).
The first instance of overlap is a function of early development and different focuses within heliophysics. 
SpacePy (see section~\ref{spacepy.sec}) was developed first and provides full CDF library support (reading and writing) along with many other tools for magnetospheric physics.
pysatCDF (see section~\ref{pysatcdf.sec}) was developed as a stand-alone python CDF reader to provide a streamlined user experience by developers in the ionospheric community.
CDFlib (see section~\ref{cdflib.sec}) was developed recently, and contains a pure Python CDF reader and writer (as opposed to a Python wrapper for the NASA CDF C library).
%
%
%
%
%
%

The second example of overlap is less avoidable.
The IGRF is used in two of the coordinate transformation packages (AACGMV2 in section~\ref{aacgmv2.sec} and pysatMagVect in section~\ref{pysatMagVect.sec}), the coordinate transformations within SpacePy, and the modeled data access and analysis package, pyglow (see section~\ref{pyglow.sec}).
The IGRF implementation within the coordinate packages and SpacePy is contained internally and it would not be desirable for them to use a common python IGRF python package.
The pyglow package is the only one of these packages to provide the user with the direct output from IGRF.
While the pyglow implementation may not be ideal for all users, since the package contains many other models and was not developed by the authors of IGRF, it does provide a unique tool needed by the heliophysics community.
%
%

Apart from obvious overlap in package functionality and model implementations, there are also instances of conceptual overlap between the packages listed in Figure~\ref{map.fig} and Table~\ref{packages.tab}.
For example, the packages that focus on the data access and analysis of observations all contain routines to read and load the observations into a python object that may be easily used for data analysis.
However, this overlap is not necessarily a problem.
Several of these packages focus on a single instrument network, allowing the experts to deal with the specific peculiarities in that data set (e.g., DaViTpy/pyDARN in section~\ref{pydarn.sec}).
Others, such as HelioPy and pysat (section~\ref{pysat.sec}), focus on the instruments in different subdisciplines of heliophysics.
The packages with broader scopes can benefit from the targeted work done by the more focused packages.
pysat sets a good example of this practice; it uses pysatCDF to read CDF files, DaViTpy/pyDARN to read Super Dual Auroral Radar Network (SuperDARN) data, pysatMagVect for magnetic vector transformations, pyglow for model access,  SGP4 and pyEphem for satellite orbit propagation, madrigalWeb for downloading data from Madrigal, as well as AACGMv2 and apexpy for magnetic coordinates. 
%
%
%
%

\section{Towards a Heliophysics Framework}
\label{framework.sec}

The broad scope of heliophysics, the diverse nature of heliophysics data sets, the lack of funding for scientific software development, and the current tradition of individualism in data analysis all present challenges for the development of a useful Python toolkit for space physics.  Despite these challenges, the community has expressed a desire to create a unified framework, similar to Astropy (described in section~\ref{astropy.sec}), for space science Python tools \citep{agu:2017}.  As the community works towards this unified framework, there are steps that can be taken to reduce duplicated efforts and increase the utility of existing packages.  These steps include establishing a steering committee, centralizing information about all Python heliophysics projects (as discussed in section~\ref{cent.sec}), providing guidelines for making these projects accessible and connectible (as described in section~\ref{access.sec}), ensuring that scientists receive appropriate recognition for their work (as discussed in section~\ref{attrib.sec}), encouraging best practices for scientists and developers (as discussed in section~\ref{best.sec}), and holding scientific software to the same standards that scientific theses are held (as discussed in section~\ref{rigor.sec}).  
Engaging in these practices will improve the trustworthiness of scientific analysis \citep{Miller:2018ni,Kanewala2014} and address many of the concerns raised by opponents of the open source philosophy described in section~\ref{benefits.sec}.

\subsection{Centralization}
%
%
%
%
\label{cent.sec}
As with many international, interdisciplinary, collaborative efforts the best way to centralize is online.  
To this end, a website has been set up to act as a hub for heliophysics Python packages at \url{http://www.heliopython.org}.  
Currently, this website focuses on providing a place to easily locate heliophysics Python packages and community developers.
Python packages are not hosted at this website, allowing developers to determine which hosting service best serves their own needs.  
To be included on this website, package authors submit a pull request on the website's GitHub page and follow the guidelines (which currently include ensuring that the software is citable and FOSS).  The code should also provide a useful tool for heliophysics research, falling into one of the categories outlined in Figure~\ref{map.fig}.
Other websites also attempt to curate subsets of heliophysics software (not limited to Python packages), including the CEDAR wiki (\url{http://cedarweb.vsp.ucar.edu/wiki/index.php/Community:Software}).

Another goal of the website is to involve new community members in development.  
By laying out where work has already been done, new developers can dedicate their time to areas in need of improvement, rather than re-inventing the wheel.  
Increasing community involvement will also help maintain active development, keep packages up to date, and improve communication between scientists involved with model development, instrument development, data analysis, and space weather products.

These community organization efforts are recent, and could benefit from following the example of more mature scientific community organizations, such as the Incorporated Research Institutions for Seismology (IRIS).  
The IRIS consortium aims to advance discovery, research, and education in seismology and has a large membership from research groups across the globe.  
Their website provides access to data, derived data products, sponsored and community developed software, education resources, archives of posters and presentations, annual reports, and much more (\url{https://www.iris.edu/hq/}).
The scope of involvement at IRIS is larger than that currently envisioned by the heliophysics Python community, but following their example and involving national and international professional organizations would increase community involvement, encourage collaboration, and reduce instances of overlapping software development.

\subsection{Accessibility}
\label{access.sec}
Software licenses are used to set the terms on which the software may be used, modified, or distributed.  
There are three types of software licenses: FOSS (which may be further divided into permissive and copyleft), proprietary, and hybrid \citep{Morin:2012hr}.  
FOSS licenses are typically used by academic institutions, since they best espouse the scientific ethos by fostering collaboration, improving reproducibility, and aiding the peer-review process.

Permissive FOSS licenses ensure the widest possible distribution and adaptation of the software, place the fewest restrictions on users, and ease the incorporation of the code by others.  
The two most common permissive FOSS licenses are the modified BSD (Berkeley Software Development) license \citep{BSD} and the MIT (Massachusetts Institute of Technology) license \citep{MIT}.  
These licenses allow commercial use, distribution, modification, and private use as long as the original developers are attributed and are not held liable (there is no warranty).  Table~\ref{packages.tab} shows that 22 of the 27 packages have a permissive FOSS license (PSF, the Python Software Foundation license is a BSD-style permissive FOSS license).

Copyleft licenses were created to assure that the benefits of FOSS are maintained in all future derivatives of the software.
The GNU (GNU's Not Unix!) General Public License (GPL) \citep{GPL} is the best-known, strong copyleft license.
These licenses explicitly describe the treatment of patents related to the software and require that any computer code linked to the licensed software have the same license.
The remaining packages in Table~\ref{packages.tab} have either a GPL or Lesser GPL (LGPL) license.

When choosing a software license, it is important to involve your local technology transfer office.  
However, for new and existing heliophysics projects to eventually form a cohesive framework, they must all have compatible licenses \citep{Morin:2012hr}.  
The adaptation of permissive FOSS licenses best supports this goal, and so should be strongly considered.

Another important aspect of accessibility is making the source code publicly available \citep{Shamir2013}.
The centralization efforts described in section~\ref{cent.sec} can only succeed if the community makes their code available online.
Few projects have the funds to set up and maintain an individual repository in perpetuity. 
Code sharing directories such as GitHub (\url{https://github.com}) and RunMyCode.org \citep{Stodden:2012} allow scientists to share, archive, and distribute software at no cost, while also providing a platform for collaborative work. 
%
%
%
%
%
%

Sometimes the analysis code for a particular project may not have a scope large enough for its authors to justify creating a repository and DOI for it.  
In such cases, it may be possible to publish source code as supplemental information in the article that shares the results from that project.  
This use of supplemental information is explicitly encouraged by some space physics journals, including the American Geophysical Union journals \citep{agujournals}.  
However, other journals (including European Geophysical Union journals) require that computer program code be deposited in a repository with a persistent identifier such as a DOI \citep{egujournals}.

\subsection{Attribution}
\label{attrib.sec}


A common argument against open source data and software is the danger of plagiarism.  When dealing with Python packages, there are two aspects of attribution that should be considered: citations and collaborations.  Citations deal with providing scientists with credit for their products (whether they be data sets, software, or theories) and collaborations deal with the ethics of building upon another person's work.

Citations are at the core of academic culture.  
Most heliophysics journals now require citations to data sets and software, discouraging plagiarism and encouraging communalism.
Properly citing FOSS used to obtain and analyze data provides an incentive for other scientists to provide quality FOSS.
Since the advent of the Digital Object Identifier (DOI), it is possible to cite software projects and data sets.  
One service that offers DOIs for software packages is Zenodo \citep{nielsen:2014}.
Unlike journal articles, which will not change after publication, software and data sets often have different versions.  
Zenodo deals with this challenge by allowing the developers to set up a DOI for a package in general and have separate DOIs for different versions.

Before software DOIs, it was common to provide attribution by entering into a collaboration with the person who created the data set that was used or provided the analysis software.  
This could be an active or an in-name only collaboration, where the work already done was acknowledge by offering a co-authorship.  
This is a valid method of attribution, though it should not replace proper referencing as described above.  
This type of consideration should also be extended when beginning a new heliophysics Python project.  
For example, before creating a Python package for an existing model, it is best practice to contact the model's author.  
This ensures that the resulting product is of the highest quality and makes it easier to maintain the package when future versions of the model are released.

Another option is for project authors to write a methods paper.
This follows the tradition of instrument papers of heliophysics, and has the advantage of allowing the project authors to describe the scope and purpose of their project, as well as outline plans for future development.
Like instrument papers, software papers may be published in either dedicated journals (such as the Journal of Open Source Software \citep{joss}) and those with a larger scope (such as the Journal of Geophysical Research: Space Physics \citep{agujournals}).
When using software that has a published methods papers, both the paper and the software DOI should be cited.
  
Because different people are often involved in each part of a project, this ensures that the efforts of all involved are recognized.  
For example, an article that uses apexpy should cite \citet{richmond:1995aa}, who describes the coordinate systems, \citet{Emmert:2010cba}, which outlines the computational process used to smoothly represent modified apex and quasi-dipole coordinates, and \citet{apexpy}, the Python implementation of \citet{Emmert:2010cba}'s code. 
In order to ensure that citations are correctly made, it is useful for the software developers to provide this information in much the same way that data and instrument teams currently do. 

\subsection{Best Practices}
\label{best.sec}

Space scientists historically have not received formal education in computer programming.
This deficit means that most of the people developing scientific analysis software are not aware of the best practices involved in writing code, providing documentation, and working with collaborators.
This should not be a barrier to publishing code.
If it is good enough to produce reliable scientific results it is both good enough for peer consumption and of interest to the scientific community \citep{Barnes:2010,Shamir2013}.
However, good coding practices (like a well written paper) improve both adaptation by others and reproducibility.
To encourage more scientists to contribute code to projects, this section provides an overview of the most relevant of these best practices. 

Best practices for coding include supporting the current standard version of the programming language, adhering to the style guide for the programming language, commenting, using descriptive variables, and reducing in-code duplication.
At the time of publication, Python 2 is a legacy version with limited future support \citep{pep373} and Python 3 is undergoing active development.
It is possible to write code compatible to both Python 2 and 3; this is the strategy that has been followed in most of the packages described in appendix~\ref{appendix.sec}.


Style guidelines for Python are described in Python Enhancement Proposal (PEP) 8 \citep{pep8}.  
PEPs are community proposed design documents used to inform the Python community about a new feature, process, or environment.
PEP 8 is a style guide for Python code that will help developers write programs that are suitable for community development and incorporation into other packages. 

Commenting and using descriptive variables ensure that a new user will be able to follow the coded algorithm and find references to the literature when certain methods or constants are used. 
For example, when coding up a theory or model that is published in a journal, it is useful to cite papers with page or equation numbers as comments when they are coded. 
Comments should be included for every function, defined constant, and algorithm stage in the software.
This low-level documentation prevents confusion over units, sources of empirically determined values, and reasons why a particular method was implemented.

Python provides two ways to comment code, through traditional source code comments and through docstrings.
Docstrings are literal strings that describe a Python package, function, class, or method, and occur as the first statement in one of these objects.
Well written docstrings are an important element of good Python code, since standard use dictates that they contain a summary of the object that they are contained in.
This practice allows docstrings to improve the maintainability and clarity of different routines, as well as speed up the learning curve for new users.
They are an improvement over standard comments because they are accessible at runtime, and so do not require the user to open up the source code.
The standard conventions for writing docstrings are described in PEP 257 \citep{pep257}.

Variable names are another potential source of elucidation or confusion.  
Best practice dictates using descriptive variables, or ensuring that the variable name indicates what it is.  
For example, iterative counts are typically have either single letter, lower case names (e.g., $i$, $j$, $k$) or names that reflect their purpose (e.g., $counter$, $inum$, $ion\_count$).

Documentation is very important, since it ensures that changes to the code (which will affect scientists abilities to replicate results in the future) can be easily identified. 
This is vital for reproducibility as scientists move onto other projects, making their expertise less accessible to their former collaborators and future investigators.
Comments and docstrings are examples of low-level documentation, providing details about specific parts of a software package.
However, their dispersion throughout a package, lack of cohesion, and specificity make them insufficient to serve as an overarching (or high-level) document. 
High-level documentation should provide general information about the project, an installation guide, useful references to appropriate literature, tutorials, and more.
Adding to high-level documentation is an excellent way for new users to contribute to software packages, since they have a perspective that the project developers lack.

The SunPy project is an excellent example of documented software.
They provide general information about the project, an installation guide, community and developer guidelines, and extensive downloadable tutorials (available at the URL shown in Table~\ref{packages.tab}).
Not all projects will be able to support such extensive online documentation.
Free online documentation is supported through a variety of websites, including \url{readthedocs.org}.
This website and other available tools often take advantage of internal, low-level documentation by turning all the docstrings in a package into a manual \citep{sphinx, numpydoc}, providing a structure around which more extenstive documentation may be written.
%
%
%
%


Duplication within a package, defined as the existence of code that is either identical to another section of code or has the same intent and structure, happens naturally during the development of complex code.
Often this happens unintentionally, when the developer does not yet realize how useful a particular portion of the algorithm will be.
Other times this happens intentionally, with the developer deciding that the abstraction necessary to remove the duplication will negatively impact functionality.
In general, duplication is problematic as it makes software packages harder to maintain.
Multiple instances of the same structure make algorithms more difficult to analyze and increases the likelihood of not fixing a known bug, since finding all instances of the bug will be more difficult \citep{Kanewala2014}.
Thus, best practice dictates that in-code duplication should be removed or managed through documentation.

\subsection{Rigor}
\label{rigor.sec}


Producing scientific analysis tools is an integral part of modern heliophsyics analysis, and so should be treated with the same level of skepticism as the results that it produces.
Peer-reviewing code is one way to ensure that computational methodology is treated with the same level of rigor as other scientific analysis methods.
Code review may be performed in-house, with collaborating developers checking each others code \citep{Kelly:2011}.
It may also be performed as a part of the publishing process.
Journals dedicated to publishing and reviewing software \citep<e.g.,>{joss} have review criteria that require the implementation of many of the best practices outlined above, including licensing, documentation, and attribution.  
In addition, the reviewers also ensure that the software under consideration is functional.

To be functional, software must conclusively address a demonstrated need. 
In science, finding a need for a software tool is not difficult.
However, testing the ability of the software to fill that need can be, since it requires that the developers determine the applicable range of conditions their algorithm can support and construct tests that convincingly demonstrate expected outcomes. 
Common software testing methods include unit, integration, system, and acceptance testing.

Unit tests are an application of component testing, a method that tests the smallest unit of an application.
Python has native and third-party frameworks for unit testing that allow each object in a program to be subjected to a series of tests, ensuring it behaves as expected under a variety of operating conditions.
Unit tests are most practically implemented as code is written, discouraging in-code duplication and reducing instances when analysis results are contaminated by software bugs.
The adoption of widespread unit testing promotes the use of short methods with outputs that are readily verifiable. 

Even when each individual component of a program is working well, problems may still be encountered when they are brought together.
Integration testing ensures that disparate parts of a code behave as expected when they invoke each other and pass data amongst themselves. 
The same frameworks that Python uses for unit testing may also be used for integration testing, since the difference between unit and integration testing is the scope of the tests rather then the implementation method.

System testing is a type of black-box testing that evaluates a computing system's compliance with the requirements of a software program.
This type of testing ensures that the program is behaving as expected in its current environment. 
There are many different types of system tests, some that are commonly used by scientific developers include installation testing (ensuring dependencies are installed and up-to-date) and regression testing (re-running integration and unit tests to ensure the software still performs as expected). 

Acceptance testing is testing performed by users.
This type of test can be performed locally, but it is also common to release a package as an `alpha' or `beta' version for this purpose.
Acceptance testing is common in heliophysics, where the small size of the community encourages contact between users and developers.
Repositories such as GitHub also assist with this level of testing, providing a channel for users to notify developers of problems they encounter when using a software package.

Best practice dictates unit, integration, and system tests should be automated, allowing an inexperienced user to run the software with confidence. 
There are several commercial services that will automatically test appropriately configured repositories \citep<e.g.,>{travis, appveyor}.
Many offer these services for free to smaller, FOSS projects. 
%
%
%
These tools promote efficient use of developer time and help users determine which packages are suitable for their research and development environment.

Despite the dangers of using untested software \citep<e.g.,>{Miller:2018ni}, scientific software is largely untested or under tested.  
The reasons behind this have been attributed to inherent difficulties in applying  tests to scientific software and the culture around scientific software development \citep{Kanewala2014}. 
The greatest technical challenge facing scientific software testing is the lack of an oracle (or known truth) to test against \citep{Kelly:2011}.
Techniques such as metamorphic, property based, and `golden run' testing may overcome some of these testing challenges, though more research from the software engineering community is needed \citep[and references therein]{Kanewala2014}.

%
%
%
%

\section{Summary}
The heliophysics community has expressed a desire to create a unified framework, similar to Astropy, for space science Python tools.  
This article has outlined some of the challenges facing those of us working towards this goal and presented guidelines that can ease the transition from individualistic and proprietary coding practices to a communalistic environment that takes advantage of each individual's expertise.  
Many of the existing tools have been outlined as well, and a central location for further information about these Python projects has been identified. 

Currently, much of this effort has been undertaken by individuals and small groups with little communication between the parties. 
This has led to significant overlap between certain packages.  
For instance, considering only the packages summarized in this article, there are three different CDF readers.
This overlap is to be expected at the beginning of a community development effort.
There also tends to be more duplication of effort between projects written by different heliophysics disciplines.
While combining these packages into a single framework may be desirable, it is more practical to create packages for each subdiscipline that play well together and rely on a common framework for truly universal tasks, such as access to file reading utilities and the calculation of plasma parameters. 

Following the best practices in software development and citations is an important step for creating an environment where collaborative software can flourish.
Although space scientists have not historically been educated in software development and may not be aware of the best practices to follow when developing software, these practices have immediate benefits that make their adoption advantageous.  
By adopting these practices in current and future work, the heliophysics community can reduce instances of unreproducible research.

Making science and scientific analysis open source, although not always possible due to institutional constraints, is consistent with scientific ethics and can increase the ease and quality of scientific research. 
FOSS for scientific analysis makes it possible to fully replicate past scientific analysis, reduces the monetary barrier to participation caused by propriety software, and allows all stages of scientific analysis to receive appropriate acknowledgement.
Ultimately, open source analysis  leads to an increase in diversity within the scientific community,  increased creativity, and better cited research. 

\appendix

\section{Description of Current Packages}
\label{appendix.sec}

The detailed descriptions provided here are intended to introduce the heliophysics community to the range of available tools and act as a reference for the packages whose scope are small enough that a paper dedicated to their description is not currently a feasible option. 

\subsection{Observational Data Access and Analysis}
\label{data.sec}
Experimental and observational scientists in the heliophysics community rely on ground and space-based data from a wide variety of instruments.  Each of these data sets have their own quirks and standards.  The Python packages in this section support data access and analysis for one or more observational data sets. 

\subsubsection{DaViTPy/pyDARN}
\label{pydarn.sec}
SuperDARN consists of High Frequency (HF) coherent scatter radars distributed over the northern and southern high and mid latitudes \citep{Greenwald:1995sd,  Chisham:2007hn}.  This network monitors the plasma convection over the poles through backscatter from field-aligned ionospheric irregularities, facilitates studies of magnetosphere-ionosphere interactions, and provides important ionospheric specifications.  The Python package pyDARN (currently provided as part of DaViTpy, but also being re-written as a more focused Python package) provides tools to retrieve, load, analyze, and visualize the SuperDARN backscatter \citep{sterne:davitpy}.

The core functionality of DaViTpy, pyDARN, sets out to provide the necessary tools to download, read, and plot the SuperDARN data.  Currently, there are routines to read the custom-formatted SuperDARN data files that are produced from raw data using the Radar Software Toolkit (RST)~\citep{rst}, obtain radar hardware information, download the files from one of the data mirrors (defaulting to the Virginia Tech mirror), perform some higher level processing, and create basic plots (such as range-time-intensity plots and maps of the radar fields of view).  DaViTpy also provides coordinate conversion tools, contains tools to perform coordinated studies with satellites and incoherent scatter radars, has a ray tracing tool, and provides Python implementations of several useful models.  This additional functionality is useful, but will not be included in pyDARN.  Modeling and coordinate system packages are better supported by existing packages (e.g., AACGMV2 and pyglow; described in sections~\ref{aacgmv2.sec} and \ref{pyglow.sec}, respectively), ray tracing tools have uses beyond the SuperDARN community and so are better developed as stand-alone packages, and coordinated studies may be performed through packages such as pysat (described in section~\ref{pysat.sec}), whose focus is providing a common framework for multiple data sets and has already successfully integrated DaViTpy. 

\subsubsection{digital\_rf}
\label{digitalrf.sec}
The Digital Radio Frequency (Digital RF) project established a disk storage and archival format for radio signals. 
It uses the Hierarchical Data Format (HDF)5, a software package and file format that can be read by any programming language \citep{hdf5}, to define a self-documenting file format for radio frequency data.  
The Python package, digital\_rf, contains routines for reading, writing, and processing radio frequency data using this format.
The digital\_rf project also has C and MATLAB implementations.
This package may be referenced using the URL provided in Table~\ref{packages.tab}.

\subsubsection{GeoData}
\label{geodata.sec}

GeoData is a software package implemented in Python and MATLAB, that plots and interpolates data from a variety of space physics sources.
The main goal of this package is to simplify the plotting and processing of geophysical data, specifically data provided by the CEDAR Madrigal database.
To support the data analysis, coordinate transformations between several geographic systems are provided.
The processing flow is streamlined by outputting data into HDF5 files.
This package may be referenced using the URL provided in Table~\ref{packages.tab}.

\subsubsection{HelioPy}
\label{heliopy.sec}
HelioPy \cite{heliopy_doi} is a Python library for heliospheric and planetary physics, whose primary goal is to make it easy to download and import common data sets. It uses CDFlib (see section~\ref{cdflib.sec}) to handle CDF files, and is set up to download and ingest a wide variety of solar and satellite data.

At the time of publication this included magnetometer and Solar Wind Ion Composition Spectrometer (SWICS) data from the Advanced Composition Explorer (ACE, \citet{Stone:1998ea}) spacecraft, magnetometer, Cluster Ion Spectrometry (CIS), and Plasma Electron And Current Experiment (PEACE) data from Cluster, particle and magnetic field data from Helios, International Monitoring Platform (IMP), the Magnetospheric MultiScale (MMS) mission, Ulysses, and Wind \citep{acuna:1995wi}, as well as magnetometer data from the THEMIS (Time History of Events and Macroscale Interactions during Substorms, \cite{Angelopoulos:2009ef}), ARTEMIS (Acceleration, Reconnection, Turbulence and Electrodynamics of the Moon’s Interaction with the Sun, \cite{Angelopoulos:2010kk}), Cassini \cite{Dougherty:2004eo}, and MESSENGER (MErcury Surface, Space ENvironment, GEochemistry, and Ranging, \cite{Anderson:2007it}) missions.
Sunspot numbers are also included, and there are plans to include DSCOVER (Deep Space Climate Observatory), NASA/Goddard Space Flight Center (GSFC) OMNI, Solar Orbiter \citep{solar_orbiter}, and Parker Solar Probe \citep{solar_probe} data.

As well as importing data, HelioPy builds upon the SpiceyPy package (described in section~\ref{spiceypy.sec}) to provide an accessible interface for performing orbital calculations.  It has also implemented a framework to perform transformations between some common coordinate systems.
Future goals for HelioPy involve building upon the Astropy package (described in section~\ref{astropy.sec}) to provide data with physical units attached, easy methods for transforming between a wider range of coordinate systems, and expanding methods for importing data.

\subsubsection{MadrigalWeb}
\label{madrigalweb.sec}
The CEDAR Madrigal Database is an online resource for archiving and retrieving many heliophysics data sets. 
This data can be accessed remotely using Python scripts, through functions provided by the MadrigalWeb package.
MadrigalWeb allows users to explore the available experiments and instruments, download the data in several file formats, calculate a range of derived parameters, and perform some instrument-specific coordinate conversions.
This package may be referenced using the URL provided in Table~\ref{packages.tab}.

\subsubsection{pysat}
\label{pysat.sec}
The Python Satellite Data Analysis Toolkit (pysat) \cite{stonebackpysat, pysat} is a high-level package intended to form a common ground for all packages and data sources in space science. 
To make this possible, pysat hides the tedious file and data handling behind a single consistent object interface in a class object called `Instrument'.
The Instrument object features robust data and metadata handling, generalized data iteration, on-the-fly orbit breakdown, and a versatile system for modifying data. 
These features enable the creation of instrument independent routines that can operate on the varied dimensionality found across space science. 

Pysat is currently being used as a framework for processing the Ion Velocity Measurements (IVM) for the upcoming NASA Ionospheric Connection (ICON) Explorer satellite \citep{Immel:2017bn} as well as the National Oceanic and Atmospheric Administration (NOAA) Formosa Satellite (Formosat)-7/Constellation Observing System for Meteorology, Ionosphere, and Climate (COSMIC)-2 Constellation. 
Several instruments from the Communication/Navigation Outage Forecasting System (C/NOFS) satellite (the IVM, Vector Electric Field Instrument (VEFI), and Planar Langmuir Probe (PLP)) \citep{delaBeaujardiere:2004uk}, NASA/GSFC OMNI data, SuperDARN grid data \citep{rst}, SuperMAG magnetometer data and indices \citep{Gjerloev:2009hq}, Formosat-3/COSMIC Global Positioning System (GPS) occultation data \citep{Liou:2007ro}, the Republic of China Satellite (ROCSAT)-1/Formosat-1 IVM \citep{su:1999rs}, Dst, Kp, the Defense Meteorological Satellite Program (DMSP) IVM \citep{Gorney:1987cc,Sun:2018kn}, the Floating Point Measurement Unit (FPMU) on the International Space Station (ISS) \citep{barjatya:2009is}, and Thermosphere Ionosphere Mesosphere Energetics Dynamics – Solar Extreme ultraviolet Experiment (TIMED-SEE) \citep{Woods:2005se} are also currently supported. 
The most recent release includes a `Constellation' class that allows simultaneous processing of heterogeneous groups of Instruments. The Constellation support was developed by undergraduate computer science students for their senior project. Upcoming versions of pysat will feature support for both Pandas \citep{mckinney-proc-scipy-2010} and xarray \citep{Hoyer:2017hs} data formats, improving support for multi-dimensioned data sets.

\subsection{Modeled Data Access and Analysis}
\label{model.sec}
The heliophysics community uses first principles and empirical models for a variety of purposes, including theoretical studies and space weather forecasts.  Modeling studies frequently face reproducibility challenges, since the data are often not made publicly available and many models are not FOSS.  The Python packages in this section begin to address these issues by providing access to documented versions of common heliophysics models, as well as standard analysis tools.

\subsubsection{analysator}
\label{analysator.sec}

Analysator is an analysis tool developed for Vlasiator, a 6-dimensional Vlasov theory-based simulation that focuses on fundamental plasma processes within the near-Earth space environment \citep{vlasiator}. 
It began as a file reader for the Vlasiator output and has evolved to include analysis and visualization tools.
Analysator facilitates studies of particle paths, pitch angle distributions, velocity distributions, and more.
More details about the capabilities of this package may be found at the URL provided in Table~\ref{packages.tab}.

\subsubsection{OvationPyme}
\label{ovationpyme.sec}
OvationPyme is a translation of the Ovation Prime model written in IDL. 
The Ovation Prime model \citep{Newell2002} predicts the total electron and ion energies and number fluxes precipitated into the upper atmosphere, as well as the characteristic energy of the precipitation (assuming a Maxwellian distribution). 
This model is based on observations from the DMSP Special Sensor Precipitating Electron and Ion Spectrometer (SSJ)4/5 particle detectors, which are sensitive to particles in the 30 eV - 30 keV energy ranges \citep{Newell:1996ea}.  
This package may be referenced using the URL provided in Table~\ref{packages.tab}.

\subsubsection{pyAMPS}
\label{pyamps.sec}
pyAMPS \citep{pyamps} is a Python interface for the Average Magnetic field and Polar current System (AMPS) model \citep{laundal:2018}. 
AMPS is an empirical model of the ionospheric current system and magnetic field, which takes inputs of the solar wind velocity, the Interplanetary Magnetic Field (IMF), the dipole tilt, and the F$_{10.7}$ index.
The primary model output is the average, large-scale, ionospheric magnetic field disturbances at any location in near-Earth space for the selected set of input parameters.
Ionospheric currents are then derived from the magnetic field. 
pyAMPS includes functions to calculate field-aligned currents, horizontal currents, and estimates of associated ground magnetic field perturbations on a grid. 
It also includes functions to calculate a time-series of model magnetic field perturbations (e.g., along satellite tracks).
The empirical model is derived from magnetic field measurements from the  Swarm \citep{Swarm} and Challenging Minisatellite Payload (CHAMP; \citet{CHAMP})  satellites. 

\subsubsection{PyForecastTools}
\label{pyforecasttools.sec}
PyForecastTools is a Python package providing implementations of a wide variety of metrics for model validation and forecast verification \citep{morley:2018pf}. 
The metrics include a generic forecast skill score, comparison metrics, scale- and order-dependent biases, a symmetric signed bias, different measures of accuracy, and common error estimates.
A key feature in this package is the inclusion of classes for contingency table analyses with multiple methods for estimating confidence intervals on scores. 
These metrics and contingency tables simplify and illuminate the model validation process, making it more accessible to new users and improving the ability of the scientific community to critically analyze model outputs and forecasts.

\subsubsection{pyglow}
\label{pyglow.sec}
Pyglow is a Python package wrapping multiple empirical Ionosphere-Thermosphere models, including the Horizontal Wind Model (HWM; \citet{Hedin:1993uh, Hedin:1993ux, Drob:2008kx, Drob:2015gj}), the International Reference Ionosphere (IRI; \citet{Bilitza:2014id, Bilitza:2017ck}), IGRF, the Naval Research Laboratory (NRL) Mass Spectrometer Incoherent Scatter radar (MSIS) Exobase (NRLMSISE)-00 model \citep{Picone:2002bc}, and an airglow model \citep{chartier:2015}.
To ensure the most up-to-date versions are used at the time of installation, pyglow retrieves the source code (for the non-Python model implementations) from the official distribution sites at the time of installation.
Pyglow also includes a package to download and archive the geophysical indices used to drive these models:  Ap, Kp, F$_{10.7}$, Dst, and AE.
This package may be referenced using the URL provided in Table~\ref{packages.tab}.

\subsection{Data Analysis and File Routines}
\label{analysis_files.sec}
Sometimes data analysis methods and file formats reach beyond disciplines.  The Python packages described in this section are used by the heliophysics community.  They have kept their scope small, though, to better serve multiple scientific fields.

\subsubsection{CDFlib}
\label{cdflib.sec}
CDFlib is a Python package for reading and writing CDF files. Unlike other CDF file handling packages, CDFlib is a pure Python implementation that does not require any compiled C or Fortran code.  This makes installing the package on different operating systems and platforms very easy.
This package may be referenced using the URL provided in Table~\ref{packages.tab}.

\subsubsection{pyLTR}
\label{pyltr.sec}
pyLTR is a Python learning-to-rank (LTR) toolkit.
LTR is a supervised machine learning method that trains a model to rank lists of data points rather than score single data points.  This is useful for applications such as information retrieval and data mining \citep{Li:2011ltr}.
 pyLTR provides ranking models, evaluation metrics, tools to load and sort data, and other relevant utilities.  pyLTR's goal is to be as full-featured as the prevailing open-source LTR library, RankLib (implemented in Java; \citet{Arnold:2000:JPL:556709}), while being significantly easier to use. 
 As such, it contains built-in support for data libraries like NumPy and Pandas.
 This package may be referenced using the URL provided in Table~\ref{packages.tab}.
 
\subsubsection{pysatCDF}
\label{pysatcdf.sec}
pysatCDF \cite{pysatCDF} provides a Python interface to the NASA CDF C library through an intermediate Fortran layer. 
To enable ease of access, pysatCDF includes the NASA library and couples the build system for the CDF C library with Python installation tools. 
This makes pysatCDF a self-contained package that is easy to install. 
pysatCDF also supports the same data access mechanisms present in SpacePy's CDF routines to enable cross-package interoperability (SpacePy is described in section~\ref{spacepy.sec}). 
pysatCDF operates independently of pysat (described in section~\ref{pysat.sec}), though it also features routines designed to simplify the integration of science data sets into pysat. 
For example, the NASA CDAWeb hosted mission support within pysat relies upon pysatCDF.

\subsection{Coordinates}
\label{coords.sec}
Coordinate systems are used to order data in a sensible fashion.  In complex and coupled systems, there is often not a single best way to do this.  As a result, there are a plethora of coordinate systems used within heliophysics.  This section outlines several packages that focus on calculating coordinate transformations between  geographic and geomagnetic systems.

\subsubsection{AACGMV2}
\label{aacgmv2.sec}
Corrected geomagnetic (CGM) coordinates are defined in terms of the intersection between the local IGRF field line and the dipole equatorial plane. The CGM longitude is the centered dipole longitude of this intersection [see \citet{laundal17} for a review]. CGM latitude is the latitude where a dipole field line with this radius intersects a sphere at $1~R_\oplus$ (Earth radius). This coordinate conversion requires magnetic field line tracing with the IGRF, and is thus relatively computation heavy. Early implementations, based on look-up tables, only allowed for conversions of points at ground. When this limitation was later removed, CGM coordinates became better known as Altitude-Adjusted Corrected Geomagnetic coordinates (AACGM). 

AACGM coordinates were originally developed for the purpose of comparing ground-based radar backscatter measurements from high-latitude locations in both hemispheres.   Originally known as the Polar Anglo-American Conjugate Experiment (PACE) geomagnetic (PGM) coordinate system \citep{Baker:1989gi}, AACGM coordinates preserve latitude and longitude along  magnetic field lines (as specified by IGRF).  AACGMv2 is the most recent incarnation of this coordinate system, providing coefficients that can be used to obtain AACGM coordinates between 0-2000 km above the surface of the earth and field line tracing for higher altitudes.  These coordinates are designed to be highly accurate at polar and middle latitudes, and may be undefined at the dip equator and near the South Atlantic Anomaly \citep{Shepherd:2014wu}.  The Python implementation of AACGMV2 provides an interface for the C code developed by \citet{Shepherd:2014wu}.  It allows conversions between geographic (or geodetic) coordinates and AACGM latitude, longitude, and local time, and it is possible to provide alternative versions of the AACGM coefficients \citep{aacgmv2}.

\subsubsection{apexpy}
\label{apexpy.sec}
Magnetic apex coordinates are defined in a similar way as CGM/AACGM coordinates. Instead of using the intersection of the IGRF model with the dipole equatorial plane, magnetic apex coordinates are based on the field line apex, the highest point above the geoid. Apex longitude is defined as the centered dipole longitude of the field line apex. Two different definitions are used for the latitude, which separate two variants of the apex coordinates: modified apex coordinates, and quasi-dipole coordinates \citep{richmond:1995aa}. 

In modified apex coordinates, the latitude is defined as the latitude where a dipole field line with radius equal to the apex height intersects a sphere with radius $R_\oplus + h_R$, where $h_R$ is a chosen reference height. In quasi-dipole coordinates, this sphere is replaced by $R_\oplus + h$, where $h$ is the height of the point of interest.  This means that modified apex coordinates are constant along IGRF magnetic field lines, while quasi-dipole coordinates are not.  If $h_R$ or $h$ are small, the apex coordinates are very similar to AACGM coordinates at high latitudes. In contrast to AACGM coordinates, apex coordinates are defined at low latitudes (above $h_R$ in the case of modified apex coordinates).

apexpy \citep{apexpy} provides functions to convert to and from apex coordinates. It also includes functions to calculate base vectors, needed to do vector calculus in these nonorthogonal coordinate systems \citep{richmond:1995aa,laundal17}, and map electric fields along magnetic field lines to different altitudes.  At apexpy's core is a wrapper for the Fortran library described in \citet{Emmert:2010cba}.

\subsubsection{OCBpy}
\label{ocbpy.sec}
High latitude ionospheric processes interact directly with the magnetosphere and the solar wind.  These interactions lead to different ionospheric behaviors in the auroral oval and the polar cap, a region where the magnetic field lines are open (connecting to the IMF).  \citet{Chisham:2017cd} developed a coordinate system for high latitudes that arranges observations relative to the Open Closed field line Boundary (OCB).  This was shown to affect the formation of empirical models and statistical studies, which could otherwise inadvertently combine observations taken in the auroral oval and polar cap.  OCBpy is a Python package that determines the location of data relative to the OCB from AACGM coordinates and, when appropriate, scales the measurements to reflect the influence of the cross-polar cap potential drop \citep{OCBpy}.  

The coordinate transformation OCBpy performs requires the OCB location, the data location, the data value, and knowledge of how this value is related to the ionospheric electric field.  OCB locations are currently provided for the northern hemisphere between May 2000 and August 2002 using observations from the Far Ultraviolet Imager (FUV) on board the IMAGE satellite \citep{Mende:2009vp, Mende:2000ct}.  Future developments will increase the number of OCB locations, incorporate AACGMV2 (described in section~\ref{aacgmv2.sec}) to allow conversion between geographic and OCB coordinates, and incorporate the pysat Instrument class (described in section~\ref{pysat.sec}) to allow coordinate transformations for a wide range of data sets.

\subsubsection{pysatMagVect}
\label{pysatMagVect.sec}
The motion of plasma in the ionosphere is constrained by the anisotropic conductivity of magnetized plasma, making perpendicular motion much more difficult than motion along magnetic field lines.
To best reflect a geophysical basis for interpreting electric fields and plasma motion, pysatMagVect calculates unit vectors in the magnetic field-aligned, meridional, and zonal directions. 
The field-aligned direction points along the magnetic field, defined to be positive when directed from south to north. 
The meridional unit vector is perpendicular to the field-aligned direction and constrained to the meridional plane, the plane that contains the field line. 
At the magnetic equator, the meridional vector is vertical and defined to be positive when directed upward. 
The zonal direction completes the orthogonal set and is positive when directed towards the East. 

The vector system is calculated by field-line tracing and vector math. Reference code for IGRF in Fortran is coupled with SciPy's Ordinary Differential Equation numerical integrator to provide a robust and accurate field-line tracing system.  
The relative locations of the field line footprints in the northern and southern hemispheres are compared to a specified location and used to define the magnetic meridian plane. 
The geomagnetic vector system may also be determined using the local magnetic field. 

Under the common assumption that geomagnetic lines are equipotentials in the ionosphere, electric fields are `mapped' along field lines.
This allows measurements made anywhere along a field line to be translated to another location on the field. 
The electric field values are not strictly maintained along the field line as the magnetic flux density changes with position. 
pysatMagVect \citep{pysatMagVect} uses field-line tracing to determine both the geomagnetic unit vectors as well as scalars needed to determine the changes in an electric field along a geomagnetic field line. 

To support these and other calculations, pysatMagVect also includes coordinate and vector transformations. 
Translation between Earth Centered Earth Fixed (where $x$ lies in equatorial plane pointing from center of Earth towards 0$^
\circ$ longitude, $y$ similarly points towards 90$^\circ$ longitude, and $z$ completes the system and is aligned with the rotation axis), as well as geographic and geodetic (WGS84) systems. 
Vector projections onto an ECEF or other custom basis specified in ECEF are also supported.

\subsection{Orbits}
\label{orbits.sec}
Space-based data from satellites and rockets requires knowledge of orbital mechanics to properly determine locations.  
The ephemera that contain orbital information are often difficult to read. 
This process becomes increasingly complex when data from multiple satellites are used together.  
The Python packages in this section provide tools for determining the orbital mechanics of natural and artificial satellites and can be very useful for mission planning, conjunction studies, and post multi-point analysis. 

\subsubsection{SpiceyPy}
 \label{spiceypy.sec}
 SPICE is a geometry information system designed, built, and maintained  by the Navigation and Ancillary Information Facility (NAIF), acting under the directions of NASA's Planetary Science Division, to assist NASA scientists in planning and interpreting scientific observations from space-borne instruments.  
 As FOSS, SPICE has been used internationally to assist spacecraft mission concept development, data analysis, and the correlation of instruments on multiple spacecraft.  
 SPICE components, or kernels, have different functions that give the system its name.  \textit{S} stands for spacecraft ephemeris, \textit{P} stands for planet, satellite, comet, asteroid, or any other target body ephemerids, \textit{I} stands for instrument information, \textit{C} stands for C-matrix (a matrix containing orientation information), and \textit{E} stands for events information (which summarizes mission activities).
 SPICE was originally implemented in Fortran 77, but is now officially supported in C, IDL, MATLAB, and Java.  
 
 There are several unofficial Python implementations of SPICE, one of which is SpiceyPy \citep{spiceypy}.  
 SpiceyPy provides a Python interface for more than 98\% of C SPICE functions, a greater percentage than are made available through officially supported IDL and MATLAB interfaces, and is thoroughly tested through the use of continuous integration services.  
 SpiceyPy also simplifies Python to C interactions by presenting an interface that simplifies C function parameters (such as array lengths and temporary data structures) to idiomatic Python and through data type conversions of common NumPy data types \citep{annex2017spiceypy}.

\subsubsection{PyEphem}
\label{pyephem.sec}
PyEphem is a Python package based on the XEphem C package \citep{xephem} that allows users to determine the position of astronomical bodies or artificial satellites with user provided orbital elements \citep{pyephem}.
These locations are available in equatorial, ecliptic, and galactic coordinates, as well as location in the sky relative to known `landmarks', such as a constellation.
Although a database of heliophysics spacecraft are not provided by PyEphem, they do provide orbital elements for a wide variety of astronomical object and long-duration artificial satellites.

\subsubsection{SGP4}
\label{sgp4.sec}
The Simplified General Perturbations \#4 (SGP4) Python package uses two line element (TLE) data for an Earth orbiting satellite to calculate its position and velocity.
The purpose of the original C++ and Python implementations of SGP4 is to foster collaboration between partners and allies by providing a high quality, FOSS propagator compatible with data products produced by the United States Air Force Space Command Joint Space Operations Center.
The Python package was  regularly tested against the C++ 2010  SGP4 propagator suit to ensure that the predictions agree within  0.1 mm. 
This package may be referenced using the URL provided in Table~\ref{packages.tab}.

\subsubsection{skyfield and jplephem}
\label{skyfield.sec}
Skyfield and jplephem are both pure Python ephemeris packages. Jplephem uses the Jet Propulsion Laboratory (JPL) ephemeris to predict  the position and velocity of a planet or other Solar System body, producing plain three-dimensional vectors.
Skyfield builds upon jplephem, computing positions for stars, planets, and Earth-orbiting satellites in a variety of coordinate systems. 
Its results are tested against the Astronomical Almanac (produced by the United States Naval Observatory and Her Majesty's Nautical Almanac Office) to within 0.5 milliarcseconds.
Each of these packages may be referenced using the URLs provided in Table~\ref{packages.tab}.

\subsection{Multipurpose}
\label{multi.sec}

With appropriate scope, vision, and resources, it is possible to provide a maintainable Python toolkit that encompasses multiple functions. 
The software outlined in this section contain functionality that crosses the categories outlined in sections~\ref{data.sec}-\ref{orbits.sec}.

\subsubsection{Astropy}
\label{astropy.sec}
The Astropy Project is a community effort to develop a core Python package for astronomy, as well as improving the usability, interoperability, and collaboration between other astronomical Python packages \citep{astropy:2013}.  
To this end, Astropy consists of a core package aimed at professional astronomers and astrophysicists and secondary packages that may or may not have been created by the core development team.
The secondary (or affiliated) packages share the goals of Astropy, and often use the core package as a base. 

The core Astropy package contains data structures and transformations, file handling, remote communication, computational tools, analysis utilities, and other supporting tools.
The data structures contain common astronomical constants, units, and coordinates that are often useful for heliophysics research.
The file handling, computational, and analysis tools are commonly used within the planetary portion of the heliophysics community, which often relies on telescope observations to observe auroral and other ionospheric emissions.

Astropy actively supports community development through affiliated packages with the aim of improving the availability of interpretable tools in the astronomical community.
A template for affiliated packages is provided to encourage new developers.
A complete list of affiliated Astropy packages is available at \url{http://affiliated.astropy.org/}.

\subsubsection{geospacepy}
\label{geospacepy.sec}

Geospacepy is a small library of Python functions used for space science data analysis, and can currently be referenced using the URL provided in Table~\ref{packages.tab}.
It includes a set of utilities for handling different time formats, some coordinate conversions, and plotting utilities for several standard data plots.
It also downloads and reads in OMNI data.
Geospacepy uses SpacePy (described in section~\ref{spacepy.sec}) to read  CDF files and PyEphem (described in section~\ref{pyephem.sec}) for astrodynamical calculations.

\subsubsection{SpacePy}
\label{spacepy.sec}
SpacePy is a Python package that contains a set of common analysis software primarily developed for the magnetospheric community \citep{SpacePy,spacepy11}. 
SpacePy includes tools to read from a variety of data formats including NASA CDF. 
A core feature is the `datamodel' module, which provides classes that allow both data and metadata to be loaded and stored in a common form from a range of sources including CDF, HDF5, and netCDF files, with full support for writing to any of these formats.
Additional functionality includes conversion between different time systems, and an interface to the International Radiation Belt Environment Modeling (IRBEM) library \citep{irbem} for computing terrestrial magnetic coordinate transformations, evaluating model magnetic fields, and tracing magnetic field lines and drift shells. 
SpacePy also provides a range of empirical models, statistical analysis tools, data handling tools, and plotting convenience functions. 

A suite of tools to work with simulation output from components of the Space Weather Modeling Framework is included in the pybats module. 
Supported components include the Block-Adaptive-Tree Solar-Wind Roe-Type Upwind Scheme (BATS-R-US;  \citet{POWELL:1999ba, dezeeuw2000}), Polar Wind Outflow Model (PWOM; \citet{glocer2009}), Ridley Ionosphere Model \citep{ridley2004}, Rice Convection Model (RCM; \citet{toffoletto2003}), Ring current Atmosphere interactions Model with Self-Consistent B field (RAM-SCB; \citet{jordanova2012}), and Global Ionosphere-Thermosphere Model (GITM; \citet{ridley2006}).

SpacePy's core development team are continuing work to improve ease of installation, ease of use, and compatibility with other packages across heliophysics. 
The SpacePy project is planning a move to an environment inspired by `scikits'  \citep{scikits}, where a streamlined core SpacePy package provides key functionality for a broad range of applications and is supplemented by specialized packages that build on the core of SpacePy.

\subsubsection{SunPy}
\label{sunpy.sec}

SunPy is a Python package for solar physics that was developed with the help and support of the global community \citep{sunpy}.
It provides a comprehensive data analysis environment that allows researchers to carry out their tasks with minimal effort.
As a mature solar physics package, SunPy handles data acquisition (from observations and models), analysis, and plotting.
It includes functions that perform common solar coordinate transformations, unit conversions, and deals with temporal parsing.
SunPy also allows local user customization of the analysis environment.

\subsubsection{PlasmaPy}
\label{plasmapy.sec}

PlasmaPy is a community developed and driven Python package for plasma physics \citep{plasmapy}.  
Still at an early stage of development, it aims to provide the common tools used within the field of plasma physics for theoretical and experimental analysis.  
It currently includes access to particle data, functions to calculate plasma parameters, dielectric tensor components, and transport coefficients.

\acknowledgments
The authors would like to acknowledge the support of all the heliophysics scientists who have participated in the CEDAR Snakes on a Spaceship workshops, the recent Snakes on a Spaceship survey, AGU meet-ups, and other community organization efforts.  We acknowledge use of NASA/GSFC’s Space Physics Data Facility's OMNIWeb service and OMNI data by many of these Python packages (\url{https://omniweb.gsfc.nasa.gov/}).  The Python packages and data accessed by these packages may be obtained by following the links provided in Table~\ref{packages.tab}.  A.G. Burrell is supported by the Chief of Naval Research.  D. Stansby is supported by the U.K. Science and Technology Facilities Council studentship ST/N504336/1.  Contributions by S.K. Morley were performed under the auspices of the U.S. Department of Energy and partly funded by the Laboratory Directed Research and Development program (grant 20170047DR).


%
\bibliography{snakesbib}
%




\end{document}